\documentclass[10pt,twocolumn,letterpaper]{article}

\usepackage{iccv}
\usepackage{times}
\usepackage{epsfig}
\usepackage{graphicx}
\usepackage{amsmath}
\usepackage{amssymb}

\usepackage{bm}
\usepackage{amsthm}
\usepackage{amsfonts}
\usepackage{bbm}
\newcommand\bmt[1]{\bm{\tilde{#1}}}
\usepackage{multirow}
\usepackage{tikz}
\usetikzlibrary{chains,scopes,positioning,backgrounds,shapes,fit,shadows,calc,arrows.meta}
\usepackage{subcaption}
\usepackage[pagebackref=true,breaklinks=true,letterpaper=true,colorlinks,bookmarks=false]{hyperref}

\iccvfinalcopy 

\def\iccvPaperID{6551} 

\ificcvfinal\fi

\begin{document}

\title{Efficient Learned Lossless JPEG Recompression}
\newcommand{\authorinstitude}[1]{{\textsuperscript{#1}}}
\author{Lina Guo\authorinstitude{1}\thanks{Equal contribution.}\ ,
Yuanyuan Wang\authorinstitude{1}\footnotemark[1]\ ,
Tongda Xu\authorinstitude{2},
Jixiang Luo\authorinstitude{1},
Dailan He\authorinstitude{1},
Zhenjun Ji\authorinstitude{1}, 
Shanshan Wang\authorinstitude{1},\\
Yan Wang\authorinstitude{2}\thanks{Corresponding author.} , 
Hongwei Qin\authorinstitude{1}\footnotemark[2] \\  
SenseTime Research\authorinstitude{1}, AIR, Tsinghua University\authorinstitude{2} \\
{\tt\small \{guolina1,wangyuanyuan,luojixiang,hedailan,jizhenjun,wangshanshan,qinhongwei\}@sensetime.com} \\
{\tt\small \{xutongda, wangyan\}@air.tsinghua.edu.cn}
}

\maketitle

\begin{abstract}
   JPEG is one of the most popular image compression methods. It is beneficial to compress those existing JPEG files without introducing additional distortion.
   In this paper, we propose a deep learning based method to further compress JPEG images losslessly.
   Specifically, we propose a Multi-Level Parallel Conditional Modeling (ML-PCM) architecture, which enables parallel decoding in different granularities. First, luma and chroma are processed independently to allow parallel coding. Second, we propose pipeline parallel context model (PPCM) and compressed checkerboard context model (CCCM) for the effective conditional modeling and efficient decoding within luma and chroma components. Our method has much lower latency while achieves better compression ratio compared with previous SOTA. After proper software optimization, we can obtain a good throughput of 57 FPS for 1080P images on NVIDIA T4 GPU. Furthermore, combined with quantization, our approach can also act as a lossy JPEG codec which has obvious advantage over SOTA lossy compression methods in high bit rate (bpp$>0.9$).
\end{abstract}

\section{Introduction}
\label{sec:intro}
\begin{figure}[t]
  \centering
  \includegraphics[width=8.3cm]{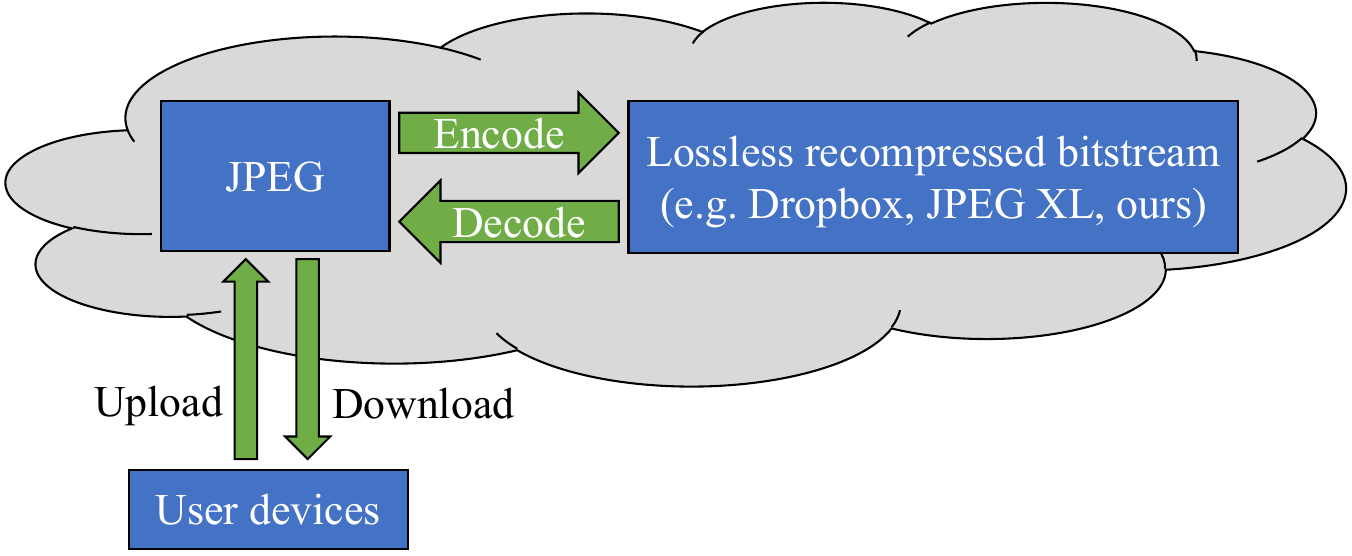}
  \caption{Lossless JPEG recompression is already adopted in real world (\eg Dropbox Lepton~\cite{horn2017design}), saving storage without affecting the compatibility of JPEG on user side.}
  \label{fig:app}
\end{figure}

Learned image compression has achieved remarkable progress in recent years and already outperforms existing traditional methods including JPEG \cite{wallace1992jpeg}, JPEG2000 \cite{rabbani2002jpeg2000}, JPEG XL \cite{ alakuijala2020benchmarking, alakuijala2019jpeg}, BPG \cite{bellard2015bpg}, and even the latest intra coding of VVC/H.266 \cite{ohm2018versatile} by a large margin. However, JPEG is still the most popular compression technique because of its simplicity and flexibility. W3Techs \cite{w3techs} find that 75.2\% of websites worldwide use JPEG image format as of July 2022. It is reasonable to 
worry that these JPEG images are not efficiently compressed, either because they are processed by discrete cosine transform (DCT) \cite{ahmed1974discrete} followed by quantization which is hard to eliminate data redundancy adequately, or because JPEG uses Huffman code \cite{huffman1952method} whose theoretical compression bound is inferior to arithmetic code \cite{10.1145/214762.214771} and Asymmetric Numeral Systems (ANS) \cite{duda2013asymmetric, duda2015use}. One solution is to further compress these existing JPEG images losslessly, namely JPEG recompression (Figure~\ref{fig:app}).

Several classical methods can be used to lossless compress JPEG images, \eg Lepton \cite{horn2017design} (already applied in Dropbox), JPEG XL \cite{ alakuijala2020benchmarking, alakuijala2019jpeg}, and PAQ8PX \cite{paq8px}, where the first two are designed for image compression while the last one for generic data compression. 

Recently, learned lossless JPEG recompression has been studied using neural networks. Guo~\etal~\cite{Guo_2022_CVPR} applies learning-based entropy model that operates on DCT domain to model data distribution and  obtains about $ 30\% $ compression savings. Fan~\etal~\cite{fan2022learned} utilizes learned end-to-end lossy transform coding to reduce  the redundancy of DCT coefficients in a compact representation domain and achieves about $21.49\%$ improvement. These two approaches have achieved superior compression savings on standard datasets than traditional methods, which shows the promise of learning-based JPEG recompression. 

On the other hand, there are multiple subsampled formats in YCbCr colorspace for JPEG images, such as YCbCr 4:4:4 and YCbCr 4:2:0, namely the resolution of chroma components (\ie Cb and Cr) is variable and may be equal to or $\frac{1}{4}$ of the luma component (\ie Y).
Current neural compressors,  integrating these three color components along channel dimension as input, are incapable of supporting YCbCr 4:4:4 and YCbCr 4:2:0 using the same model. 
Another limitation is that their structure is not conducive to optimize throughput, which hinders real-world applications.

\begin{figure*}[t]
  \centering
  \includegraphics[width=16cm]{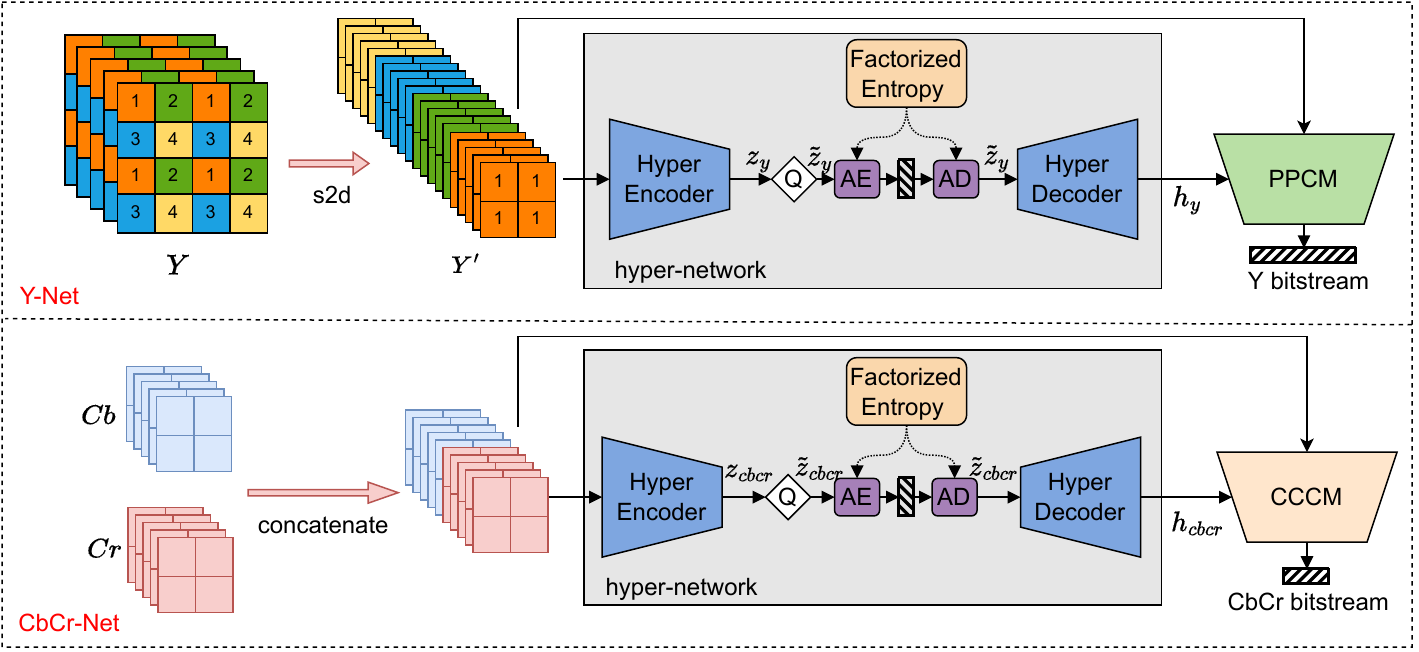}
  \caption{Overview of our Multi-Level Parallel Conditional Modeling
(ML-PCM) architecture. }
  \label{fig:overview}
\end{figure*}

In this paper, we propose a more lightweight and flexible neural compressor composed of two neural networks, \ie Y-Net and CbCr-Net in Figure~\ref{fig:overview}, where Y-Net is used to compress luma component (\ie Y) while CbCr-Net to chroma components (\ie Cb and Cr). In this case, our compressor not only is compatible with a variety of JPEG formats, but also has higher computation efficiency due to luma and chroma being coded in parallel. We design two efficient context models for luma and chroma components respectively, namely pipeline parallel context model (PPCM) and  compressed checkerboard context model (CCCM). 



In conclusion, our contributions include:
\begin{itemize}
    \item We propose a novel Multi-Level Parallel Conditional Modeling (ML-PCM) architecture 
    for lossless JPEG recompression, which is compatible with various JPEG formats and enables parallel decoding in different granularities. 
    \item Experiments show that ML-PCM achieves state-of-the-art performance on benchmark datasets with faster running speed (Table~\ref{tab:comparison:set} and Table~\ref{tab:latency}). 
    \item ML-PCM can be extended to JPEG lossy compression and achieves significantly better RD performance than previous SOTA lossy compression method under higher bit rates (Figure~\ref{fig:comparsion_lossy}). 
    \item ML-PCM is designed to be friendly to multi-threading and streaming (Figure~\ref{fig:flops}), reaching astonishing throughput after proper software optimization (Table~\ref{tab:throughput}), which can provide new insight for efficient learned image or video compression.
\end{itemize}

\section{Related work}
\subsection{Overview of JPEG compression}
\label{ssec:jpeg}
JPEG first converts image from RGB sources to YCbCr colorspace and then selects a format for subsampling, including YCbCr 4:4:4, YCbCr 4:2:2, YCbCr 4:2:0 and so on. Subsequently, each component is divided into multiple $8\times8$ blocks and each block is transformed by discrete cosine transform (DCT) into frequency coefficients (\ie DCT coeficients). Specially, the coefficient with zero frequency is 
called DC coefficient while the remaining 63 coefficients are called AC coefficients. Next, these three color components are quantized by quantization tables to filter out information insensitive for human visual system, especially high frequencies and color details stored in chroma components. Finally, quantized coefficients are compressed by Huffman code with a probability model defined by Huffman tables.

\label{sec:relate}
\subsection{JPEG recompression methods}

\textbf{Traditional.} 
There are some traditional methods devoted to recompress JPEG images. Mozjpeg \cite{mozjpeg_code, mozjpeg} improves the encoder of JPEG to achieve smaller file size while maintaining compatibility with those already deployed JPEG decoders. It losslessly reduces file size by 10\% on average for a sample of 1500 JPEG images from Wikimedia. 

In addition, general data compression programs can also further compress JPEG images, such as PAQ8 \cite{paq} and CMIX \cite{cmix}. 
PAQ8 is a series of compressor archivers, including famous PAQ8PX \cite{paq8px}. In fact, PAQ8PX is also adopted in CMIX. Though these compressors are different, they all employ a key technique called context mixing, where massive context models independently predict the next bit of input and then pick the most precise prediction for current step. Therefore, these methods achieve higher savings (about 23\%), while they are considerably slower and consume more computation and memory.

Recently, more practical traditional compressors have emerged, \eg Lepton \cite{horn2017design} and JPEG XL \cite{ alakuijala2020benchmarking, alakuijala2019jpeg}. Lepton replaces Huffman code in JPEG with more efficient arithmetic code \cite{10.1145/214762.214771} and uses a sophisticated adaptive probability model which produces more accurate predictions. It achieves about 22\% compression savings after recompressing JPEG losslessly. 
JPEG XL achieves further compression of JPEG file by extending the $8\times8$ DCT to variable-size DCT, \eg allowing block size to be one of 8, 16 or 32. And it uses ANS \cite{duda2015use} in place of Huffman code.  

\textbf{Learning-based}
Guo~\etal~\cite{Guo_2022_CVPR} directly learns a probability model by learning-based entropy model on DCT domain, which efficiently reduces the mismatch between estimated data distribution and true distribution. They achieve about 30\% savings and outperform traditional compressors by a large margin. Fan~\etal~\cite{fan2022learned} points out that there exists considerable redundancy among DCT coefficients because discrete cosine transform is unable to eliminate redundant data adequately. They transform DCT coefficients into a compact representation by learned end-to-end lossy transform, and then code this representation and the residual between lossy recovery  and original coefficients. They achieve about $21.49\%$ savings over JPEG.

However, these two methods use integrated color components as input and are incapable of supporting different JPEG subsample formats with single model. And their latency and throughput are not good enough for practical applications. Our multi-level parallel design is more effective and efficient, and is more compatible with software optimization techniques like multi-threading and streaming.

\subsection{End-to-end lossless image compression}
Actually, learned lossless JPEG recompression is a special use case of end-to-end lossless image compression. In theory, data can be compressed into bitstreams losslessly by entropy code with a probability model. According to Shannon’s source coding theory \cite{Shannon1948A}, the lower bound of the bitstream length is limited by the entropy of data’s ground truth distribution. However, the true distribution is unkown, entropy coder is generally applied with estimated distribution. Mismatch between the approximate distribution and the true distribution will bring overhead, \ie the preciser the probability model is, the shorter the bitstream will be. 

Recently, deep generative models (DGMs) have shown the powerful ability of approximating distribution. It is not surprising that many compression algorithms based on DGMs have emerged and obtain the recent state-of-the-art performance in terms of compression ratio. These algorithms can be roughly categorized into four groups:
autoregressive model \cite{child2019generating,HawthorneJCBNMD22,oord2016conditional,parmar2018image,reed2017parallel,roy2021efficient,van2016pixel,ZhangZM21a}
, variational autoencoder 
\cite{flamich2020compressing,kang2022pilc,kingma2019bit,mentzer2019practical,ryder2022split,townsend2018practical,townsend2019hilloc}
, flow model 
\cite{ho2019compression,hoogeboom2019integer,van2020idf++,wang2022fast,zhang2021ivpf}
, and diffusion model \cite{barbu2019segmentation,hoogeboom2021autoregressive,kingma2021variational}. 
Except for DGMs, some neural lossless compressor \cite{cao2020lossless,rhee2022lc,zhang2020lossless} uses context based entropy model for distribution approximation, which essentially is the improved variant of the autoregressive model in computational complexity. Moverover, some methods learn lossless compressor by compressing the residual of lossy compressors, such as \cite{bai2021learning,mentzer2020learning}.


Those lossless compressors all operate on RGB domain and are designed to compress images stored in PNG format. According to Guo~\etal~\cite{Guo_2022_CVPR}, these methods are not effective when used to losslessly compress JPEG images. We provide an analysis to show that DCT domain is indeed preferred for JPEG lossless recompression.
\section{Method}
\label{sec:method}
\subsection{Overview}
We apply the same data processing as Guo~\etal~\cite{Guo_2022_CVPR} to rearrange each color component from $(h\times 8,  w\times 8)$ to $(h, w, 64)$.
To support multiple subsampled formats, we design two independent networks for chroma components and luma component respectively in Figure~\ref{fig:overview}. Each network is composed of a \textit{hyper-network} (\ie hyper encoder and hyper decoder) and a parallel \textit{context model} (\ie CCCM or PPCM), where \textit{hyper-network} extracts side information to learn global correlation while context model captures more local details from decoded adjacent symbols to further reduce redundancy. The DCT coefficients of Y component is first transformed into $\bm{Y}'$ by space-to-depth (s2d) and then compressed by Y-Net (top of Figure~\ref{fig:overview}). Meanwhile, the DCT coefficients of Cb and Cr components are concatenated at channel dimension and compressed by CbCr-Net (bottom of Figure~\ref{fig:overview}). Specifically, 
the \textit{hyper-network} in Y-Net or CbCr-Net is simple and has similar structure as the hyperprior network in previous lossy compression methods \cite{minnen2018joint} (detailed in the appendix), while CCCM and PPCM are novel and specially designed according to the traits of different color components, which is detailed in Section~\ref{ssec:arc}. 

\subsection{Why DCT domain is preferred?}
Guo~\etal~\cite{Guo_2022_CVPR} empirically shows that DCT domain brings superior performance than pixel domain. In this paper, we provide theoretical justification under mild assumption. Denote the image in pixel domain as $\bm{x}$, and its DCT transform coefficient as $\bm{w}=A\bm{x}$, and the quantized coefficient with quantization step-size $\Delta$ as $\bmt{w}=\Delta \lfloor\bm{w}/\Delta \rceil$, where $\lfloor.\rceil$ is the rounding operator. We follow the $\rho$-domain assumption, which is commonly adopted in video coding \cite{10.1117/12.453117}. More specifically, we assume that the bitrate to encode quantized symbol is proportional to the number of non-zero dimension after quantization. More formally, denote $R(\bmt{w})$ as the bitrate to encode quantized symbol $\bmt{w}$, we have:
\vspace{-0.5em}
\begin{equation}
    \vspace{-0.5em}
    R(\bmt{w}) \propto (1-\rho(\bmt{w})) \textrm{, where } \rho(\bmt{w}) = \sum_{i=1}^{D} \mathbbm{1}_{\bmt{w}^i=0}
\end{equation}

Denote the inversely transformed pixel domain symbols as $\bmt{x} = A^{-1}\bmt{w}, $we have the following property:
\vspace{-0.5em}
\begin{align}
\vspace{-0.5em}
    \rho(\bmt{x}) = \sum_{i=1}^D \mathbbm{1}_{\bmt{x}^i=0} = \sum_{i=1}^D \mathbbm{1}_{\sum_{j=1}^D A^{ij}\bmt{w}^j=0}
\end{align}

It is obvious that $\rho(\bmt{x}) < \rho(\bmt{w})$, as 
$p(\sum_{j=1}^D A^{ij}\bmt{w}^j=0)$
is very small, unless for very special $A$ such as identity matrix. Then, it becomes obvious that $R(\bmt{w})<R(\bmt{x})$, given the original quantization happens on $\bm{w}$ instead of $\bm{x}$. If we assume this conclusion on CABAC~\cite{1218195} (an ad-hoc entropy model) also applies to our entropy model, then we can say that the DCT-domain coding is preferred.

\begin{figure}[t]
   \centering
   \begin{subfigure}{0.5\linewidth}
      \centering
      \includegraphics[height = 2cm]{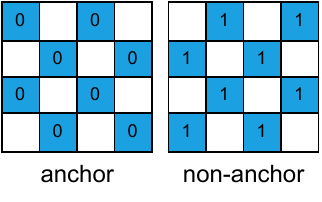}
      \caption{Checkerboard~\cite{he2021checkerboard}}
      \label{fig:compare:ckbd}
  \end{subfigure}
  \begin{subfigure}{0.49\linewidth}
      \centering
      \includegraphics[height = 2cm]{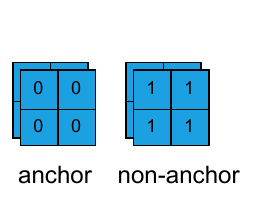}
      \caption{CCCM}
      \label{fig:compare:cccm}
  \end{subfigure}
  \caption{CCCM vs Checkerboard.}
  \label{fig:compare_ckbd}
\end{figure}

\begin{figure}[htb]
  \centering
  \includegraphics[width=8.3cm]{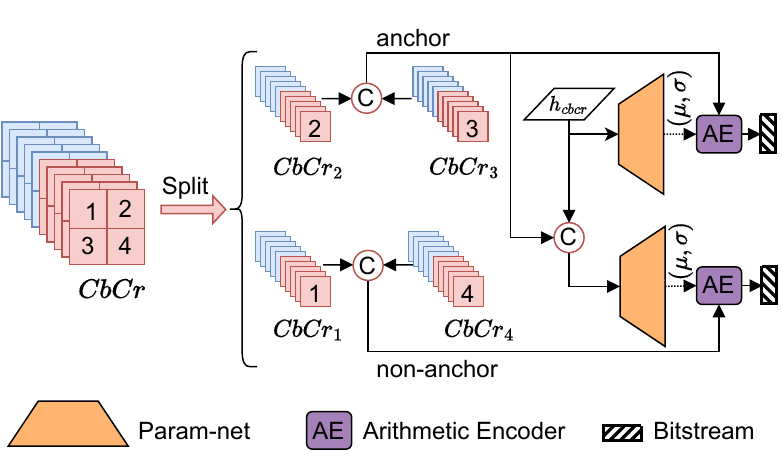}
  \caption{Compressing CbCr with CCCM and $\bm{h}_{cbcr}$. \textit{Param-net} is detailed in the appendix.}
  \label{fig:ccm}
\end{figure}

\begin{figure*}[htb]
  \centering
  \includegraphics[width=16cm]{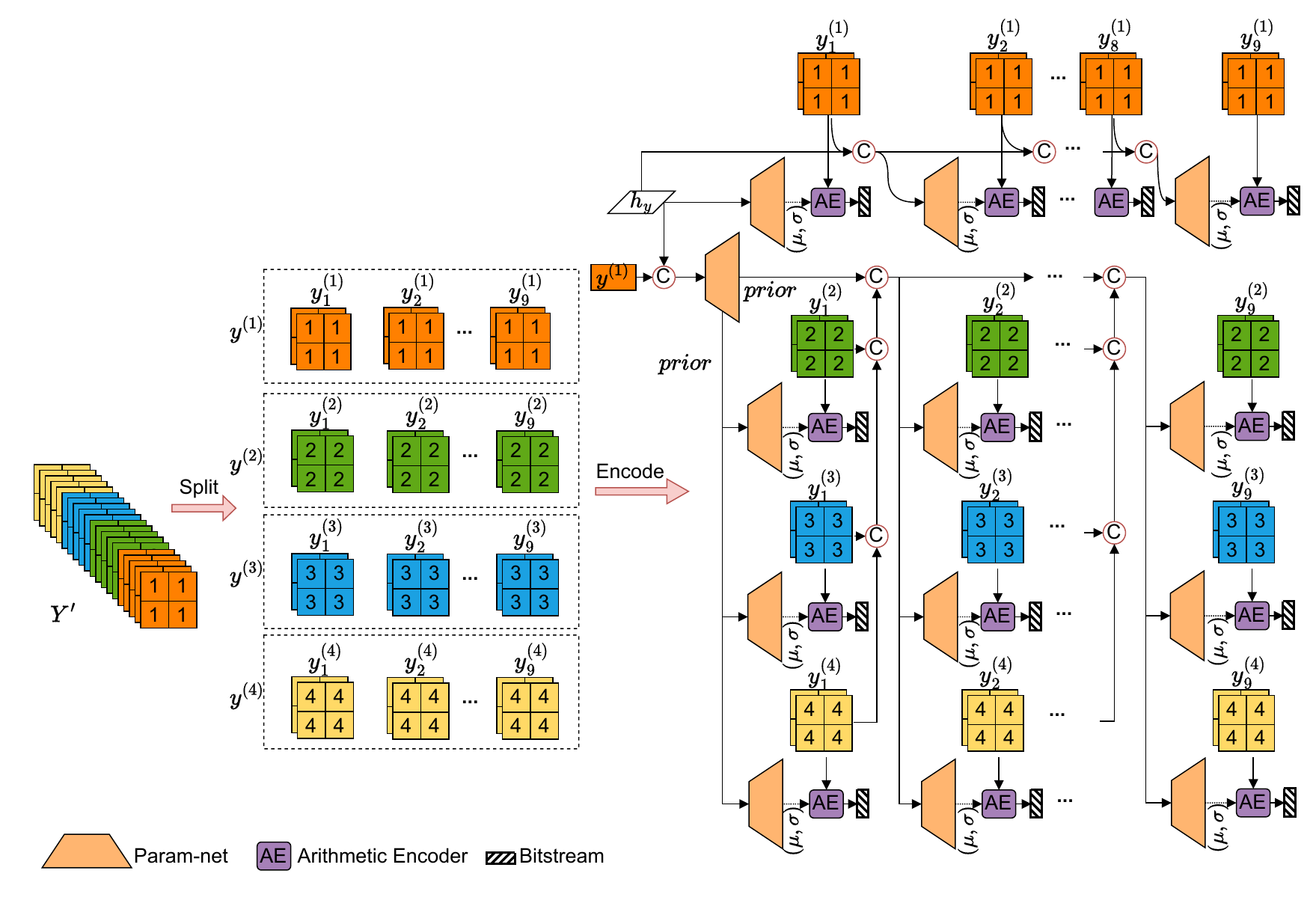}
  \caption{Compressing Y with PPCM and $\bm{h}_y$. \textit{Param-net} is detailed in the appendix.}
  \label{fig:ppcm}
\end{figure*}

\subsection{Architecture}
\label{ssec:arc}

\textbf{CCCM.}
\label{ssec:ckbd}
He \etal \cite{he2021checkerboard} proposes a parallel context model with a two-pass decoding approach, where the symbol tensor is first decomposed into two groups (\ie anchor and non-anchor in Figure~\ref{fig:compare:ckbd}) by checkerboard-shaped mask, and then anchor serves as context of non-anchor to build conditional distribution. Although they eliminate the limitation of serial context model and speed up the decoding process, some redundant computation is involved because anchor and non-anchor use the same resolution as the symbol tensor. Therefore, we propose a compressed checkerboard context model (CCCM) for CbCr-Net by categorizing groups in a more compact way (Figure~\ref{fig:compare:cccm}). As shown in Figure~\ref{fig:ccm},  concatenated tensor of Cb and Cr is grouped into four groups $ \bm{CbCr}_i $ ($i \in \left\{1,2,3,4 \right\}$, denoting the location). Then, $ \bm{CbCr}_2 $ and $ \bm{CbCr}_3 $ are concatenated in the channel dimension  as anchor , $ \bm{CbCr}_1 $ and $ \bm{CbCr}_4 $ as non-anchor. The anchor is compressed using a single Gaussian entropy model with mean and scale conditioned only on $\bm{h}_{cbcr}$ which is obtained by $\bmt{z}_{cbcr}$ through the \textit{Hyper Decoder}, while the entropy model for the non-anchor is conditioned on both $\bm{h}_{cbcr}$ and anchor.



\textbf{PPCM.}
\label{ssec:ppcm}
We propose a pipeline parallel context model (PPCM) to compress Y component, which enables more parallelism and less latency.
As shown in Figure~\ref{fig:ppcm}, $\bm{Y}'$ (obtained by space-to-depth on $\bm{Y}$ ) is first split into matrix representation $\bm{y}^{(i)}_{j}$, 
$ i \in \left\{1,2,3,4 \right\}$ denotes row index, $j \in \left\{1,2,\dots,9 \right\} $ denotes column index, and the lengths of each column $\bm{y}^{(i)}_{j}$ are set as 28, 8, 7, 6, 5, 4, 3, 2 and 1 respectively. Next, PPCM learns more powerful probability mass function (PMF) for each column using a single Gaussian entropy model with mean and scale conditioned on context and $ \bm{h}_y$. Specially, the first block $\bm{y}^{(1)}_{1}$ is predicted only by $ \bm{h}_y$ from the \textit{Hyper Decoder}, while the entropy parameters for remaining columns in the first row (\ie $\bm{y}^{(1)}_{j}, j=2,3,\cdots,9$) are conditioned on all the decoded coefficients in previous columns ($ \ie  \left\lbrace \bm{y}^{(1)}_{1}, \cdots, \bm{y}^{(1)}_{j-1} \right\rbrace $). The PMF for all columns in first row can be formulated as:
\begin{equation}
	\label{eq:row1}
	\begin{aligned}
		 &p\left(\bm{y}^{(1)}|\bmt{z}\right)   = 
		 \prod^{9}_{j=1} p\left(\bm{y}^{(1)}_{j}|\bm{C}^{(1)}\right) \\
		 &p\left(\bm{y}^{(1)}_{j}|\bm{C}^{(1)}\right) = \prod^{m_{j}}_{k=1} p\left(  \bm{y}^{(1)}_{jk}|\bm{C}^{(1)}\right)\\
		 &p\left(\bm{y}^{(1)}_{jk}|\bm{C}^{(1)}\right)= \int_{\bm{y}^{(1)}_{jk}-\frac{1}{2}}^{\bm{y}^{(1)}_{jk}+\frac{1}{2}} \mathcal{N}(y'|\mu_{\bm{y}^{(1)}_{jk}}, b_{\bm{y}^{(1)}_{jk}}) \,dy'
	\end{aligned}
\end{equation}
where $\bm{C}^{(1)} = \left\lbrace \bm{y}^{(1)}_{j-1}, \cdots, \bm{y}^{(1)}_{1}, \bmt{z} \right\rbrace$ denote the context for $\bm{y}^{(1)}_j$, $ \bm{y}^{(1)}_{jk} $ is coefficient $ k $ in column $ j $ at first row $\bm{y}^{(1)}$, $ m_{j} $ is the number of coefficients in column $ j $, $ j=1, 2, \cdots, 9 $, and  $ k=1, 2, \cdots, m_{j} $.

After first row is processed, they will be concatenated with $\bm{h}_y$ and then sent to a \textit{param-net} network to acquire $\bm{prior}$ for $\bm{y}^{(2:4)}$. Unlike the strict processing order for columns in Guo~\etal~\cite{Guo_2022_CVPR}, the columns in $\bm{y}^{(2:4)}$ are conditioned on $\bm{prior}$ and previously decoded columns in a pipeline parallel manner. Specifically, entropy parameters of columns with same index $j$ (\ie $\bm{y}^{(2:4)}_{j}$) can be calculated in parallel, and after these three context elements in column $ j $ are compressed, they will be concatenated with $\bm{prior}$ and then sent to three \textit{param-net} to obtain entropy parameters for $\bm{y}^{(2:4)}_{j+1}$. Repeat this operation until all columns in $\bm{y}^{(2:4)}$ are compressed. The calculation of PMF for all columns in $\bm{y}^{(2:4)}$ can be formulated as  follows:
\vspace{-0.5em}
\begin{equation}
	\label{eq:row234}
	\begin{aligned}
			&p\left(\bm{y}^{(2:4)}|\bm{y}^{(1)},\bmt{z}\right) = 
		 \prod^{9}_{j=1} p\left(  \bm{y}^{(2:4)}_{j}|\bm{C}^{(2)}\right) \\
		 &p\left( \bm{y}^{(2:4)}_{j}|\bm{C}^{(2)}\right) \\
		 		 & = p\left(  \bm{y}^{(2)}_{j}|\bm{C}^{(2)}\right)p\left(  \bm{y}^{(3)}_{j}|\bm{C}^{(2)}\right)p\left(  \bm{y}^{(4)}_{j}|\bm{C}^{(2)}\right) \\
		 &p\left(  \bm{y}^{(i)}_{j}|\bm{C}^{(2)}\right) = \prod^{m_{j}}_{k=1} p\left(  \bm{y}^{(i)}_{jk}|\bm{C}^{(2)}\right)\\
		 &p\left(\bm{y}^{(i)}_{jk}|\bm{C}^{(2)}\right)= \int_{\bm{y}^{(i)}_{jk}-\frac{1}{2}}^{\bm{y}^{(i)}_{jk}+\frac{1}{2}} \mathcal{N}(y'|\mu_{\bm{y}^{(i)}_{jk}}, b_{\bm{y}^{(i)}_{jk}}) \,dy'
	\end{aligned}
\end{equation}
where $\bm{C}^{(2)} = \left\lbrace \bm{y}_{j-1}^{(2:4)}, \cdots, \bm{y}_1^{(2:4)}, \bm{y}^{(1)}, \bmt{z} \right\rbrace$ denote the context for $\bm{y}^{(2:4)}_j$, $ \bm{y}^{(i)}_{jk} $ is coefficient $ k $ in column $ j $ at first row $\bm{y}^{(i)}$, $ m_{j} $ is the number of coefficients in column $ j $, $ j=1, 2, \cdots, 9 $, and  $ k=1, 2, \cdots, m_{j} $.


\begin{figure}[htb]
  \centering
  \includegraphics[width=.47\textwidth]{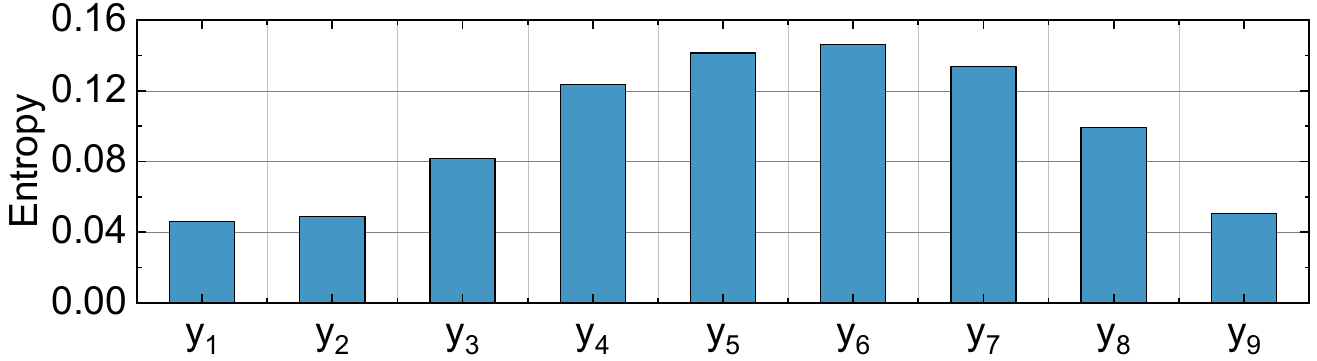}
  \caption{Entropy distribution along columns in Y-Net without shift context. $\bm{y}_j$ denotes total entropy of $\bm{y}_j^{(1:4)}$.}
  \label{fig:channel_entropy}
\end{figure}

\begin{figure}[htb]
  \centering
  \includegraphics[width=6cm]{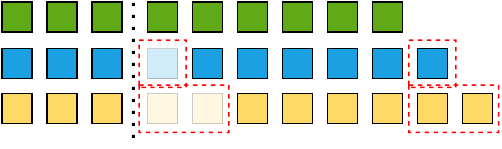}
  \caption{The shift context. }
  \label{fig:shift}
\end{figure}

\textbf{Shift Context.}
All coefficients in the same column of different rows (\eg $\bm{y}^{(1:4)}_{j}$) represent the same frequency with different space location, while different columns in same rows (\eg $\bm{y}^{(i)}_{1:9}$) represent different frequency with same space location. However, all columns in $\bm{y}^{(2:4)}$ rely on previous columns and are coded with same order, 
\ie $\bm{y}^{(2:4)}_{1}\rightarrow \bm{y}^{(2:4)}_{2}\rightarrow \cdots \rightarrow \bm{y}^{(2:4)}_{9}$, which only explores the correlation across frequency. To exploit more correlation cross spatial dimension, we shift the coding order of columns in $\bm{y}^{(3)}$ and $\bm{y}^{(4)}$ to allow a subset of coefficients to learn from the coefficients at different space location in the same frequency, which is presented in Figure~\ref{fig:shift}. 
As shown in Figure~\ref{fig:channel_entropy}, entropy of the first three columns is much lower, which indicates that these columns have already been effectively compressed. 
Therefore, we shift the order from the fourth column, \ie $\bm{y}^{(2:4)}_{1}\rightarrow\bm{y}^{(2:4)}_{2}\rightarrow \bm{y}^{(2:4)}_{3}\rightarrow \bm{y}^{(2)}_{4}, \bm{y}^{(3)}_{5}, \bm{y}^{(4)}_{6} \rightarrow \bm{y}^{(2)}_{5}, \bm{y}^{(3)}_{6}, \bm{y}^{(4)}_{7} \rightarrow \bm{y}^{(2)}_{6}, \bm{y}^{(3)}_{7}, \bm{y}^{(4)}_{8} \rightarrow \bm{y}^{(2)}_{7}, \bm{y}^{(3)}_{8}, \bm{y}^{(4)}_{9} \rightarrow \bm{y}^{(2)}_{8}, \bm{y}^{(3)}_{9}, \bm{y}^{(4)}_{4}\rightarrow \bm{y}^{(2)}_{9}, \bm{y}^{(3)}_{4}, \bm{y}^{(4)}_{5}$. Table~\ref{tab:ablation} demonstrates that our Shift Context not only reduces FLOPs but also brings higher compression savings.



\section{Experiments}
\label{sec:experiments}
\textbf{Datasets.} Following Guo~\etal~\cite{Guo_2022_CVPR}, we train our model on the largest 8000 images chosen from the ImageNet \cite{deng2009imagenet} validation set. 
We evaluate on  four datasets, Kodak \cite{kodak}, DIV2K \cite{agustsson2017ntire}, CLIC \cite{clic} professional and CLIC mobile.
We use \textit{torchjpeg.codec.quantize\_at\_quality} \cite{ehrlich2020quantization} to extract quantized DCT coefficients with given JPEG quality level, and then feed these coefficients to model.

\textbf{Implementation details.}
 Training images are cropped randomly into $ 256 \times 256 $ patches and then quantized DCT coefficients are extracted. Y-Net and CbCr-Net are implemented in PyTorch \cite{paszke2019pytorch} and optimized separately. We adopt Adam optimizer with learning-rate $ 10^{-4 }$ and batch-size $16$. We apply gradient clipping (clip\_max\_norm = 0.5)  for the sake of stability and decay learning rate to $ 10^{-5 }$ for last 500 epochs. 
 Y-Net is trained with three stages. We first train \textit{hyper-network} to fully learn the global information, then train PPCM to capture detailed context, and finally finetune them together. Table~\ref{tab:ablation} shows that Y-Net without three stage will deteriorate about 0.7\% compression savings.  


\textbf{Deployment and software optimization.} 
To better show the practicality of our method,
We additionally deploy our model with TensorRT \cite{tensorrt} library. 
We leverage 8-bit quantization \cite{esser2020learned} to further accelerate inference, which achieves comparable compression performance (only deteriorating about 1\% savings for Y-Net and 0.7\% for CbCr-Net) with the full precision model.

\begin{table*}[]
\begin{center}
\begin{tabular}{c|c|p{2.4cm}p{2.4cm}p{2.4cm}p{2.4cm}}
    \hline
    ~ &~ & \multicolumn{4}{|c}{BPP and Savings (\%)}\\
    \hline
    Source format&Method & Kodak & DIV2K  & CLIC.mobile & CLIC.pro\\
     \hline
     \multirow{4}{*}{JPEG 4:2:0}&
     JPEG \cite{wallace1992jpeg}            
     & 1.369    &   1.285  &  1.099  & 0.922\\
     ~&Lepton \cite{horn2017design}
     & 1.102 \textcolor[RGB]{0,180,0}{$ (19.50\%) $} &1.017 \textcolor[RGB]{0,180,0}{$ (20.86\%) $}& 0.863  \textcolor[RGB]{0,180,0}{$ (21.47\%) $} & 0.701  \textcolor[RGB]{0,180,0}{$ (23.97\%) $}\\
     ~&JPEG XL \cite{ alakuijala2020benchmarking, alakuijala2019jpeg}
     & 1.173 \textcolor[RGB]{0,180,0}{$ (14.30\%) $} &1.072 \textcolor[RGB]{0,180,0}{$ (16.58\%) $}& 0.908
     \textcolor[RGB]{0,180,0}{$ (17.38\%) $} & 0.744  \textcolor[RGB]{0,180,0}{$ (19.30\%) $}\\
     ~&CMIX \cite{cmix}
     & 1.054 \textcolor[RGB]{0,180,0}{$ (23.01\%) $} &0.931 \textcolor[RGB]{0,180,0}{$ (27.55\%) $}& 0.804
     \textcolor[RGB]{0,180,0}{$ (26.84\%) $} & 0.648  \textcolor[RGB]{0,180,0}{$ (29.72\%) $}\\
     ~&Guo \cite{Guo_2022_CVPR}  
     & 0.965 \textcolor[RGB]{0,180,0}{$(29.51\%)$}
     & 0.892 
     \textcolor[RGB]{0,180,0}{$(30.58\%) $} 
     & 0.772 \textcolor[RGB]{0,180,0}{$(29.75\%)$} 
     & 0.624 \textcolor[RGB]{0,180,0}{$(32.32\%)$}
     \\
     ~&Ours  & \textbf{0.959} \textcolor[RGB]{0,180,0}{\bm{$(29.96\%)$}}& \textbf{0.887} \textcolor[RGB]{0,180,0}{\bm{$(31.0\%) $}} & \textbf{0.755} \textcolor[RGB]{0,180,0}{\bm{$(31.32\%)$}} & \textbf{0.608} \textcolor[RGB]{0,180,0}{\bm{$(34.06\%)$}}\\
    \hline
             \multirow{4}{*}{JPEG 4:4:4}&
     JPEG \cite{wallace1992jpeg}        
     & 1.601     &   1.566   &  1.271  & 1.140\\
     ~&Lepton \cite{horn2017design}
     & 1.261  
     \textcolor[RGB]{0,180,0}{$ (21.27\%) $} &1.220  
     \textcolor[RGB]{0,180,0}{$ (22.10\%) $}
     & 0.964   
     \textcolor[RGB]{0,180,0}{$ (24.11\%) $} 
     & 0.844   
     \textcolor[RGB]{0,180,0}{$ (25.98\%) $}\\
     ~&JPEG XL \cite{ alakuijala2020benchmarking, alakuijala2019jpeg}
     & 1.348  
     \textcolor[RGB]{0,180,0}{$ (15.81\%) $} &1.272  
     \textcolor[RGB]{0,180,0}{$ (18.78\%) $}
     & 1.013 
     \textcolor[RGB]{0,180,0}{$ (20.28\%) $} 
     & 0.881   
     \textcolor[RGB]{0,180,0}{$ (22.70\%) $}\\
     ~&CMIX \cite{cmix}
     & 1.180 \textcolor[RGB]{0,180,0}{$ (26.3\%) $}
     &$ -- $
     & $ -- $
     & $ -- $ \\ 
     ~&Guo(dct444) \cite{Guo_2022_CVPR}  
     & 1.122 \textcolor[RGB]{0,180,0}{$(29.90\%)$}
     & 1.088
     \textcolor[RGB]{0,180,0}{$(30.54\%) $} 
     & 0.888 \textcolor[RGB]{0,180,0}{$(30.14\%)$} 
     & 0.769 \textcolor[RGB]{0,180,0}{$(32.59\%)$}\\
     ~&Ours  
     & \textbf{1.093} \textcolor[RGB]{0,180,0}{\bm{$(31.72\%)$}}
     & \textbf{1.065} \textcolor[RGB]{0,180,0}{\bm{$(32.02\%) $}} & \textbf{0.844} \textcolor[RGB]{0,180,0}{\bm{$(33.58\%)$}} & \textbf{0.735} \textcolor[RGB]{0,180,0}{\bm{$(35.57\%)$}}\\
     \hline
\end{tabular}
\end{center}
\caption{Performance comparison on various datasets. 
}
\label{tab:comparison:set}
\end{table*}

\subsection{Performance}
\label{ssec:comparison}
{ \bf Performance comparison with other JPEG recompression methods.} We compare the proposed model against other state-of-the-art methods for JPEG recompression on four test datasets mentioned above. 
Our model can not only be compatible with a variety of JPEG formats, but also achieve higher compression savings than previous SOTA. As shown in Table~\ref{tab:comparison:set}, with quality level set as 75, our method achieves lowest bit rate on all evaluation datasets and obtains around $ 30\% $ compression savings. Specially, for input in JPEG 4:4:4, CMIX is difficult to compress high resolution images (\eg DIV2K, CLIC.mobile and CLIC.pro) beacuse of its high computation complexity.

\begin{figure}[htb]
  \centering
  \begin{subfigure}[b]{.9\linewidth}
      \centering
      \includegraphics[width=.9\linewidth]{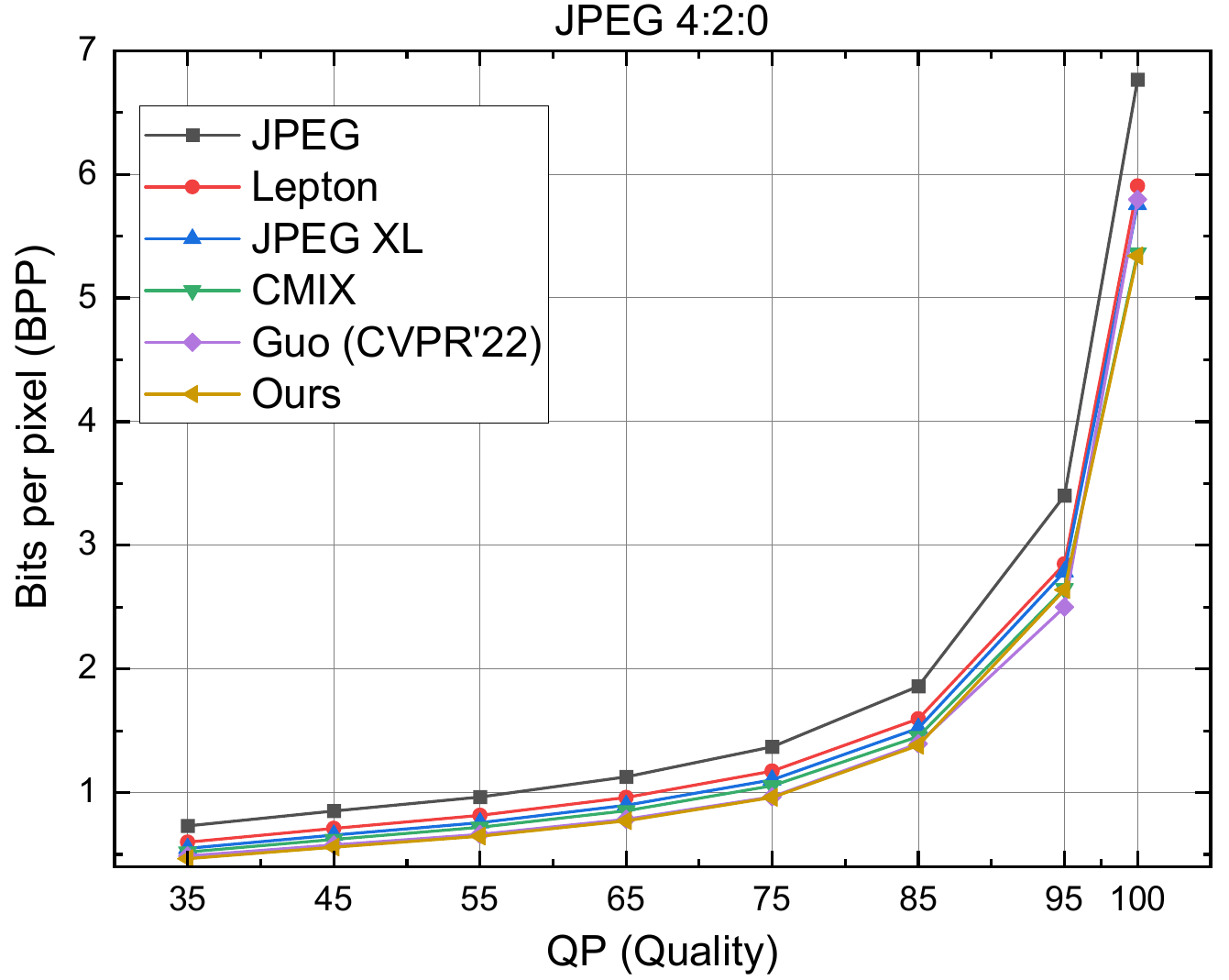}
      \label{fig:comparison:qp420}
  \end{subfigure}
  \begin{subfigure}[b]{.9\linewidth}
      \centering
      \includegraphics[width=.9 \linewidth]{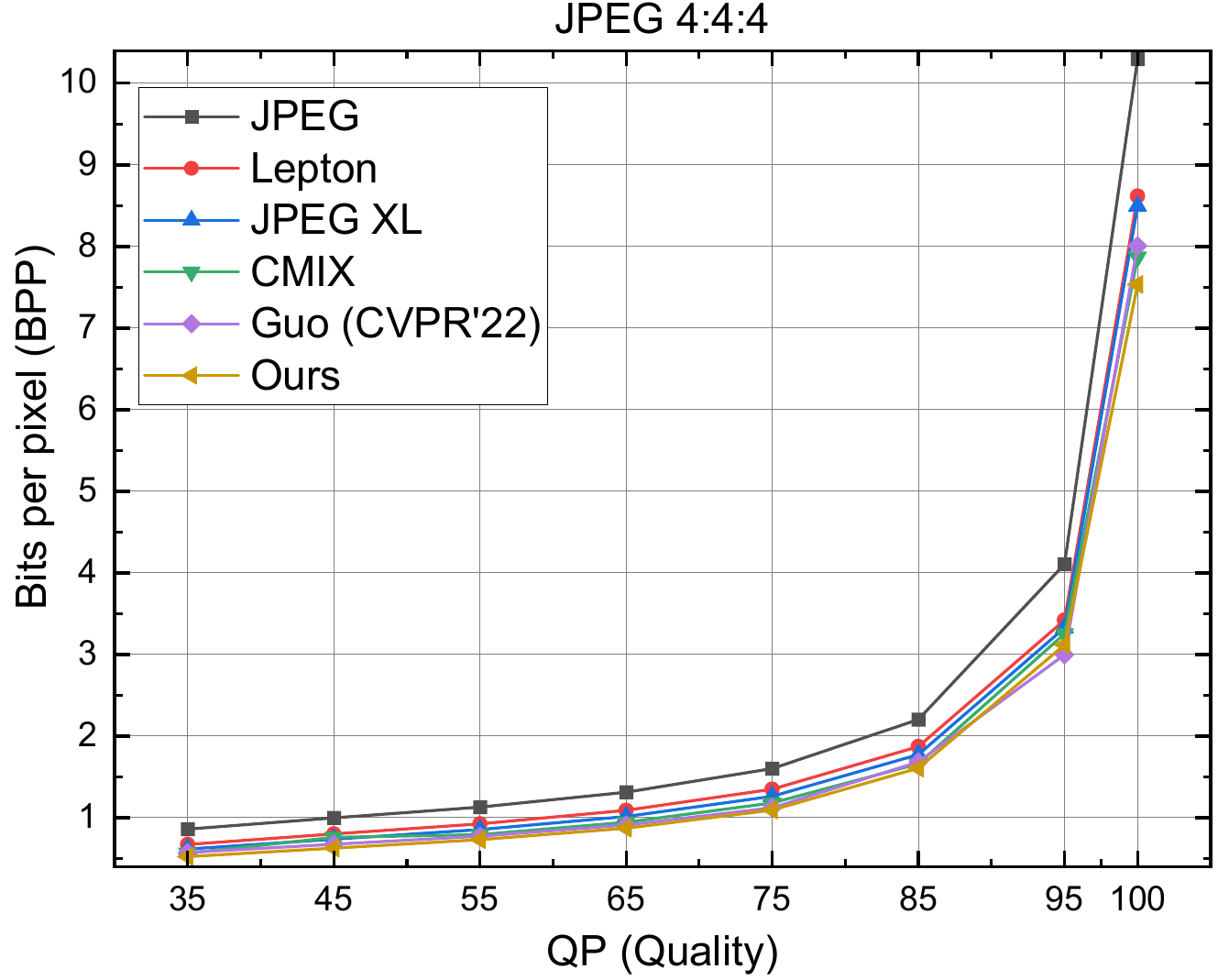}
      \label{fig:comparison:qp444}
  \end{subfigure}
  \caption{Comparison of bits per pixel (BPP) on Kodak dataset when recompressing JPEG images of different quality levels (QP). At QP$>85$, we use the model trained with mixed QP 96, 97, 98, 99 and 100. At other points, we use the model trained with QP 75.}
  \label{fig:comparison:qp}
\end{figure}

\textbf{Performance on different quality levels.}
We test our models on Kodak with two source formats at $ 8 $ different JPEG quality levels (\ie $ quality=35, 45, 55, 65, 75, 85, 95, 100 $) in Figure~\ref{fig:comparison:qp}. It shows that our method still outperforms other methods, which demonstrates our model, only using two sets of parameters, can well compress all quality levels including $QP > 95$. 

\begin{figure*}[htb]
  \centering
  \includegraphics[width=0.99\linewidth]{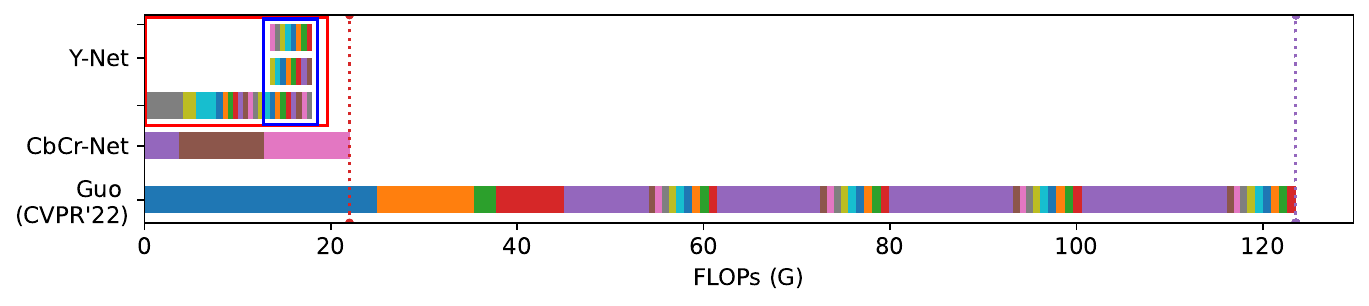}
  \caption{Comparison of network parallelism during decompressing. Horizontal bars (or rows) represent calculations executed on different pipelines, different colors represent different sub-modlues and length represents FLOPs per sub-modlue (evaluated on Kodak with JPEG 4:4:4 at QP 75). Our method is designed to be friendly to multi-threading and streaming.}
  \label{fig:flops}
\end{figure*}

\textbf{Network latency.}
Guo~\etal~\cite{Guo_2022_CVPR} evaluated network latency of their model, L3C \cite{mentzer2019practical}, IDF \cite{hoogeboom2019integer} and multi-scale \cite{zhang2020lossless}, which shows their model is much faster. Table~\ref{tab:latency} shows that our model has achieved about 40\% lower latency than Guo~\etal~\cite{Guo_2022_CVPR} on decoding, even though our PyTorch implementation doesn't do parallel optimization, \ie all sub-models are computed in serial. Actually, we can achieve even lower latency with pipeline parallelism optimization. We compare the network operation with Guo~\etal~\cite{Guo_2022_CVPR} during decompressing in Figure~\ref{fig:flops}. All FLOPs in Guo~\etal~\cite{Guo_2022_CVPR} have strict order and must be operated in serial. However, the FLOPs of Y-Net (in red box) and CbCr-Net in our model can be operated in parallel. Moreover, all FLOPs in PPCM for predicting columns $\bm{y}^{(2:4)}_j$ (in blue box) can be operated in parallel. Finally, our model's operation ends at the red dotted line, while Guo~\etal~\cite{Guo_2022_CVPR} at the purple dotted line. 

\begin{table}[]
    \centering
     \begin{tabular}{l|c|c|c}
     	\hline
         \multirow{2}{*}{Model} & \multirow{2}{*}{FLOPs} & \multicolumn{2}{c}{Latency (ms)}
         \\
         \cline{3-4}
         ~&~&Encode&Decode \\
        \hline
         Guo \etal \cite{Guo_2022_CVPR}                    
         & 136.11G& 35.30&31.64 \\
         \hline
         Ours&59.1G & 22.52&18.98\\
         \hline
         Y-Net                    
         &30.84G &15.05 & 13.73\\
         CbCr-Net                   
         &28.26G &7.46 &5.25 \\
         \hline
    \end{tabular}
    \caption{Comparison of network latency on Kodak dataset with JPEG 4:4:4 at QP 75. All results are test by PyTorch on single Nvidia GeForce GTX 1660 6GB (GPU).}
    \label{tab:latency}
\end{table}

\textbf{Results of software optimization.}
As shown in Table~\ref{tab:throughput}, with proper software optimization, our method can obtain lower latency than JPEG XL, and achieves 57 FPS for 1080P images, which shows the promise of our method. 
\begin{table}[]
    \centering
     \begin{tabular}{l|c|c|c}
     	\hline
         \multirow{2}{*}{Model} & \multirow{2}{*}{device} & \multicolumn{2}{c}{Latency (ms)}
         \\
         \cline{3-4}
         ~&~&Encode&Decode \\
        \hline
         Lepton\cite{horn2017design}                  
         & CPU& 42.06&21.62 \\
         \hline
         JPEG XL \cite{ alakuijala2020benchmarking, alakuijala2019jpeg}
         & CPU& 52.79&68.50 \\
         \hline
         CMIX \cite{cmix}
         & CPU& 4.1$ \times 10^5 $
         &4.2$ \times 10^5$ \\
         \hline
         Ours (int8)          
         & GPU\&CPU& 52&54\\
         Ours (fp16)          
         & GPU\&CPU& 53&54\\
         \hline
             	\hline
         \multirow{2}{*}{Model} & \multirow{2}{*}{device} & \multicolumn{2}{c}{Throughput (FPS)}
         \\
         \cline{3-4}
         ~&~&Encode&Decode \\
        \hline
         Lepton\cite{horn2017design}                  
         & CPU& 132.4&177.78 \\
         \hline
         JPEG XL \cite{ alakuijala2020benchmarking, alakuijala2019jpeg}
         & CPU& 103.09&141.64 \\
         \hline
         Ours (int8)          
         & GPU\&CPU& 57&57\\
         Ours (fp16)          
         & GPU\&CPU& 44.4&48\\
         \hline
    \end{tabular}
    \caption{Comparison of throughput and latency for 1080p images. CPU  is Intel(R) Core(TM) i9-10900X CPU @ 3.70GHz and GPU is T4. Lepton (without decoding validation) and JPEG XL are tested on CPU with their optimal number of threads, \ie 16 and 20 threads respectively. CMIX cannot set threads, so we use its default 1 thread. Our network inference runs on T4 GPU using TensorRT while entropy coder runs on CPU with 1 thread.}
    \label{tab:throughput}
\end{table}
\subsection{Comparison with lossy compression}
\label{ssec:lossy}

\begin{figure}[htb]
  \centering
  \includegraphics[width=7cm]{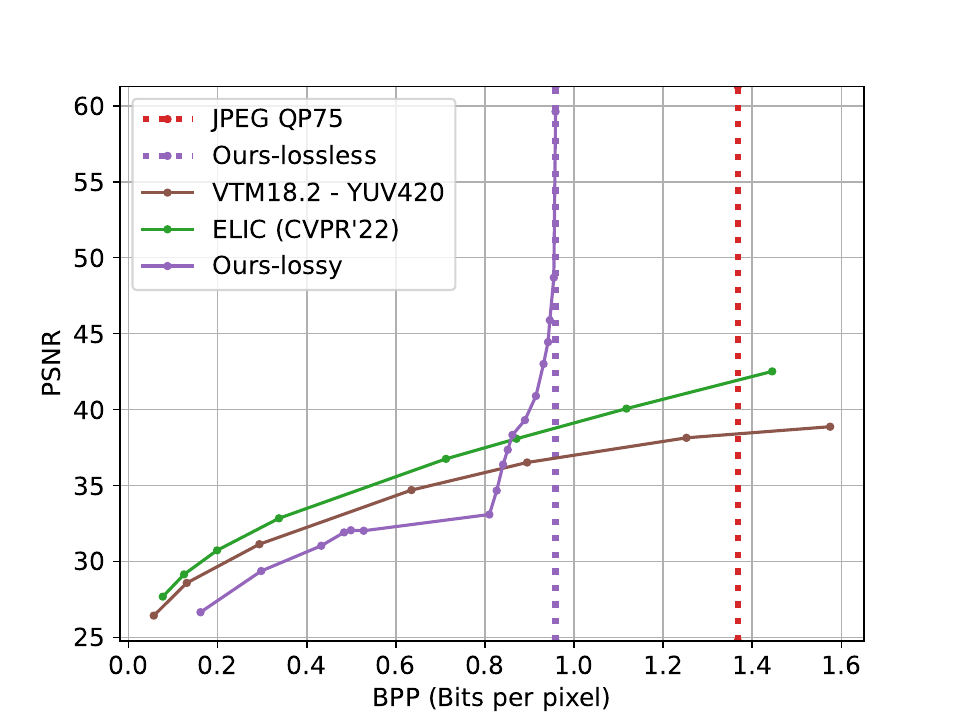}
  \caption{Comparison with lossy compression methods when taking JPEG images as input. The results are evaluated on Kodak with JPEG 4:2:0 at QP 75.}
  \label{fig:comparsion_lossy}
\end{figure}

Lossy image compression focuses on the compression of images stored in lossless format like PNG, serving as replacement of JPEG. However, JPEG algorithm is more widely used in real application. 
If lossy compression methods are used to compress JPEG images, we will find an interesting discovery shown in Figure~\ref{fig:comparsion_lossy}.
For input images in JPEG format, after the bit rate exceeds a certain threshold, both learned (ELIC~\cite{he2022elic}) and non-learned (VTM, the latest intra coding from VVC/H.266 \cite{ohm2018versatile}) lossy compression methods reach higher bit rate than JPEG lossless recompression, and even higher than original JPEG file. In other words, lossy compression methods are not effective for JPEG lossy recompression under high bit rate. 

We add some simple operations on DCT coefficients to make our method support lossy JPEG recompression without training. For example, for input DCT coefficients at QP 75, these coefficients are first dequantized, and then quantized again with quantization table at other QP, like QP 65 or QP 55 and so on. 
Then, the requantized coefficients are sent to our model and transcoded into bitstream. Finally, we achieve RD performance shown in purple solid line (marked as Ours-lossy) in Figure~\ref{fig:comparsion_lossy}, which illustrates that this simple variant of our method can achieve significantly better RD performance than previous SOTA lossy compression method for JPEG input under higher bit rates, revealing the improvement space of lossy image compression. 

\subsection{Ablation study}
\label{ssec:ablation}

\textbf{PPCM.}  \textbf{w/o ppcm} drops the PPCM in Y-Net, which means there is only hyper-network to estimate the entropy parameters for Y component. Shown in Table~\ref{tab:ablation}, without PPCM,
the bit rate of Y component increases severely (about 17\%). Besides, \textbf{w/ full parallel ppcm} adapts PPCM into a fully parallel structure (more detailed architecture in appendix): $\bm{y}^{(1:4)}_{j}$ can be coded in parallel, and then be concatenated with $\bm{h}_y$ to help predicting next group. We find this variant  deteriorates about 1.5\% compression savings. 

\begin{table}[]
    \centering
     \begin{tabular}{l|c|c|c}
     	\hline
         Method &Parameters & FLOPs & BPP \\
         \hline
         JPEG \cite{wallace1992jpeg} (Y)
         &-- &--  &1.245 \\
         Y-Net                    
         &12.05M &30.84G  &0.885 \\
         w/o ppcm
         &4.09M &6.434G  &1.102\\
         w/ full parallel ppcm
         &11.19M &28.19G  &0.905\\
         w/o shift context
         &12.08M &30.90G  &0.892\\
         w/o stage traning
         &12.05M &30.84G  &0.894 \\
         \hline
         JPEG \cite{wallace1992jpeg} (Cb\&Cr)
         &-- &--  &0.356 \\
         CbCr-Net               
         &9.23M &28.26G &0.211 \\
         w/o cccm             
         &4.20M &27.09G &0.248 \\
         w/ checkerboard             
         &4.66M &38.21G &0.211 \\
         \hline
    \end{tabular}
    \caption{Ablation study. We test a serial of models on Kodak with JPEG 4:4:4 at QP 75.}
    \label{tab:ablation}
\end{table}

\textbf{Shift Context.}
\label{ablation:shift}
\textbf{w/o shift context} removes shift context from Y-Net. As shown in Table~\ref{tab:ablation}, 
we find shift context can not only further reduce about 0.56\% bit rate , but also slightly save parameters and computation.

\textbf{CCCM.} Table~\ref{tab:ablation} shows CCCM increases calculation slightly (about 1.17 GFLOPs), but helps chroma components reduce about 10\% bit rate. We also replace CCCM with original checkerboard \cite{he2021checkerboard} and find checkerboard's complexity is higher without compression benefits.

\section{Conclusion}
\label{sec:conclusion}
We propose a novel Multi-Level Parallel Conditional Modeling (ML-PCM) architecture for lossless recompression of existing JPEG images, which is compatible with various JPEG formats and achieves SOTA compression performance. Our model achieves at least 40\% lower network latency than previous SOTA and reaches a good throughput after proper software optimization. Furthermore, for lossy JPEG recompression, a simple variant of our method obtains significantly better RD performance than previous SOTA lossy compression method in high bitrate. 

{\small
\bibliographystyle{ieee_fullname}
\bibliography{ms}
}
\end{document}


\title{Supplementary Material}


\maketitle
\ificcvfinal\thispagestyle{empty}\fi

\section{Loss function}
\label{ssec:loss}
When entropy coder is applied with our learned distribution as its probability model, the expected code length can be given by the cross entropy:
\begin{equation}
	\label{eq:loss}
	\begin{aligned}
		R  &= \mathbb{E}_{\bm{\widetilde{z}}_{cbcr} \sim  \widetilde{p}_{\bm{\widetilde{z}}_{cbcr}}}\left[ -\log_2 p_{\bm{\widetilde{z}}_{cbcr}}(\bm{\widetilde{z}}_{cbcr})\right]  \\
			 &+ \mathbb{E}_{\bm{cbcr} \sim  \widetilde{p}_{\bm{cbcr} |  \bm{\widetilde{z}}_{cbcr}}}\left[ -\log_2 p_{\bm{cbcr} |  \bm{\widetilde{z}}_{cbcr}}(\bm{cbcr} |  \bm{\widetilde{z}}_{cbcr}) \right]   \\
		     &+ \mathbb{E}_{\bm{\widetilde{z}}_{y} \sim  \widetilde{p}_{\bm{\widetilde{z}}_{y}}}\left[ -\log_2 p_{\bm{\widetilde{z}}_{y}}(\bm{\widetilde{z}}_{y})\right]  \\
			 &+ \mathbb{E}_{\bm{y} \sim  \widetilde{p}_{\bm{y} |  \bm{\widetilde{z}}_{y}}}\left[ -\log_2 p_{\bm{y} |  \bm{\widetilde{z}}_{y}}(\bm{y} |  \bm{\widetilde{z}}_{y}) \right]   \\
	\end{aligned}
\end{equation}
where $ \widetilde{p} $ is the true distribution of DCT coefficients, $ p $ is learned by our model. Our model is trained to minimize this cross entropy to  optimize the bit-length.

\section{Decoding Procedure}
\label{ssec:decode}
\subsection{Chroma component decoding}
\begin{figure}[htb]
  \centering
  \includegraphics[width=8cm]{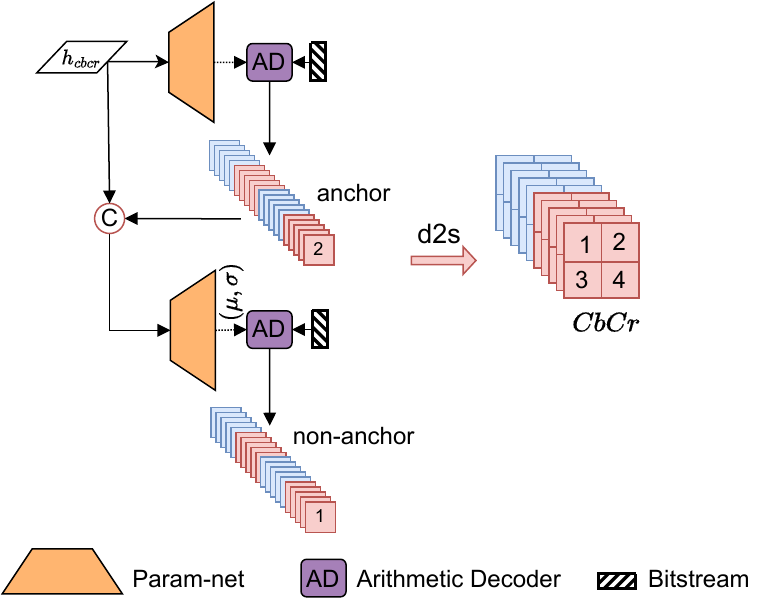}
  \caption{Decompressing bitstream to chroma component (\ie Cb and Cr) with CCCM and $\bm{h_y}$. Detailed architecture of \textit{param-net} is shown in Figure~\ref{fig:param}.}
  \label{fig:chroma_de}
\end{figure}

For the decoding, $\bmt{z}_{cbcr}$ is first decoded from bitstream using factorized entropy model, and then sent to \textit{Hyper Decoder} in CbCr-Net's \textit{hyper-network} to produce $\bmt{h}_{cbcr}$. Subsequently, chroma component (\ie Cb and Cr) are decompressed from bitstream with CCCM and $\bm{h}_{cbcr}$. As shown in in Figure~\ref{fig:chroma_de}, anchor is first decoded only with $\bm{h}_{cbcr}$. Next, anchor is combined with $\bm{h}_{cbcr}$ to decode non-anchor. Finally, anchor and non-anchor tensor are restored to chroma component by depth-to-space (d2s), where d2s is the inverse of space-to-depth (s2d) in the encoding procedure.

\subsection{Luma component decoding}
\begin{figure*}[htb]
  \centering
  \includegraphics[width=17cm]{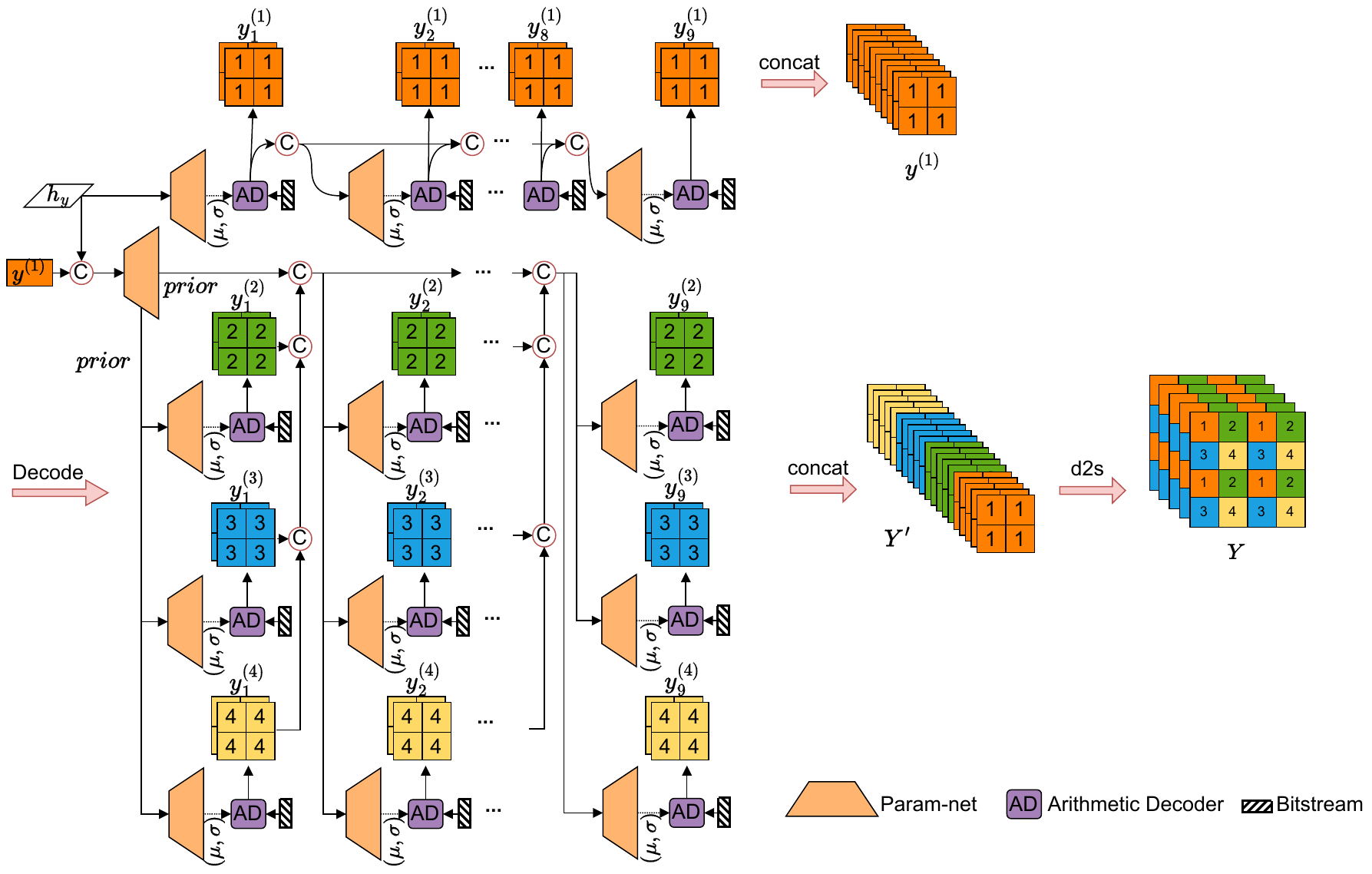}
  \caption{Decompressing Y with PPCM and $\bm{h_y}$. Detailed architecture of \textit{param-net} is shown in Figure~\ref{fig:param}.}
  \label{fig:ppcm_de}
\end{figure*}

Same as the decoding procedure of chroma component, $\bmt{z}_y$ is first deocded and then fed to \textit{Hyper Decoder} in Y-Net's \textit{hyper-network} to obtain $\bm{h}_y$. Finally, luma component (Y) is decompressed with PPCM and $\bm{h}_y$. Details of decompressing Y are illustrated in Figure~\ref{fig:ppcm_de}.

Particularly, the first block $\bm{y}_1^{(1)}$ is decoded only with $\bm{h}_y$, while the remaining columns in the first row (\ie $\bm{y}_j^{(1)}, j=2,\cdots,9$) are decoded relying on all decoded coefficients in previous columns ($ \ie  \left\lbrace \bm{y}^{(1)}_{1}, \cdots, \bm{y}^{(1)}_{j-1} \right\rbrace $).

After all columns in first row is decoded, they will be restored to $\bm{y}^{(1)}$ and then concatenated with $\bm{h}_y$ to acquire $\bm{prior}$ which is used to decode the remaining columns in $\bm{y}^{(2:4)}$. Specifically, these columns with same index $j$ (\ie $\bm{y}^{(2:4)}_{j}$) will be decompressed in parallel. And after these three columns are decompressed, they will be concatenated with $\bm{prior}$ and then sent to three \textit{param-net} to obtain entropy parameters for decoding $\bm{y}^{(2:4)}_{j+1}$. Repeat this operation until all columns in $\bm{y}^{(2:4)}$ are decompressed. Finally, all columns are concatenated as $\bm{Y}'$ and restored to $\bm{Y}$ using d2s operation.

\section{Similarity visualization analysis}
\label{sec:similarity}
We measure the similarity of DCT coefficients by cosine distance and use \textit{torch.cosine\_similarity} \cite{cosine_similarity} to acquire similarity coefficient directly. Next, the similarity is analyzed in two aspects.

\begin{figure}[htb]
  \centering
  \includegraphics[width=7cm]{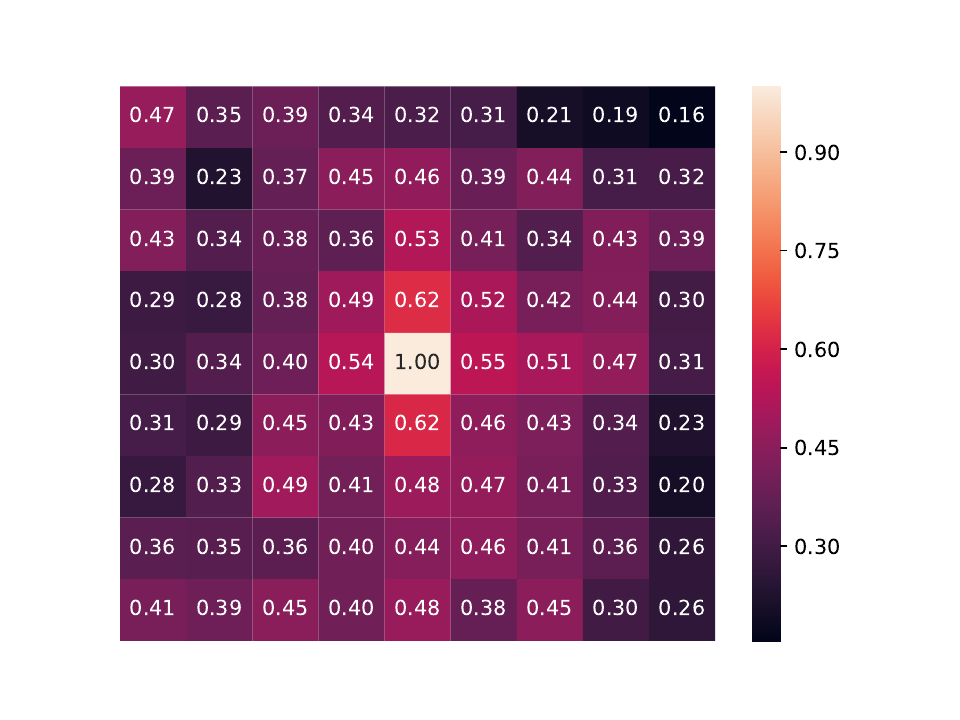}
  \caption{Similarity between per $8 \times 8$ block tested on Kodak images with JPEG 4:2:0 at QP 75.}
  \label{fig:between_blocks}
\end{figure}

\subsection{Similarity between blocks.} Every $8\times 8$ block in JPEG is viewed as a 64-dimensional vector, \eg coefficients' shape from $(h,  w, 8, 8)$ to $(h,  w, 64)$, and then the similarity measurement is made between the central block and the other blocks, which is presented in Figure~\ref{fig:between_blocks}. There are strong similarity between each $8 \times 8$ blocks, especially those blocks with nearer locations. 
It helps explain why grouping color components into $\bm{y}^{(i)}$ in PPCM or $\bm{CbCr}_i$ in CCCM (mentioned in Section 3.3 in main text) will bring compression benefits.

\begin{figure*}[htb]
  \centering
  \includegraphics[width=17cm]{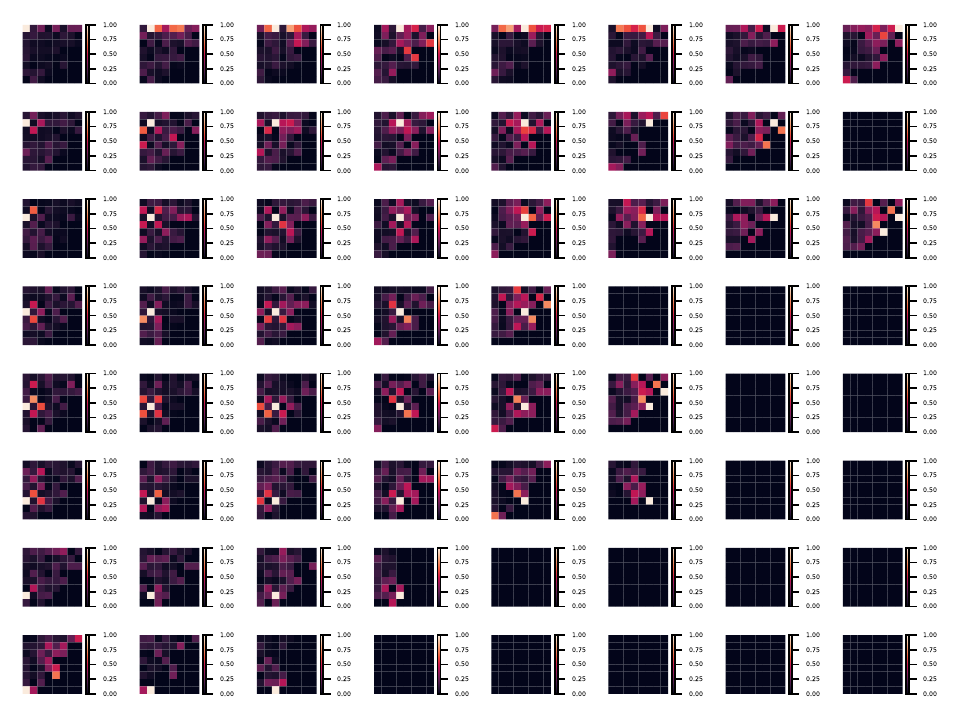}
  \caption{Similarity between different frequency coefficients tested on Kodak images with JPEG 4:2:0 at QP 75. For example, The top left corner diagram presents the similarity of the DC coefficient with 63 AC coefficients, and so on with the rest of the diagram.}
  \label{fig:in_block}
\end{figure*}

\begin{figure}[htb]
  \centering
  \includegraphics[width = .45\textwidth]{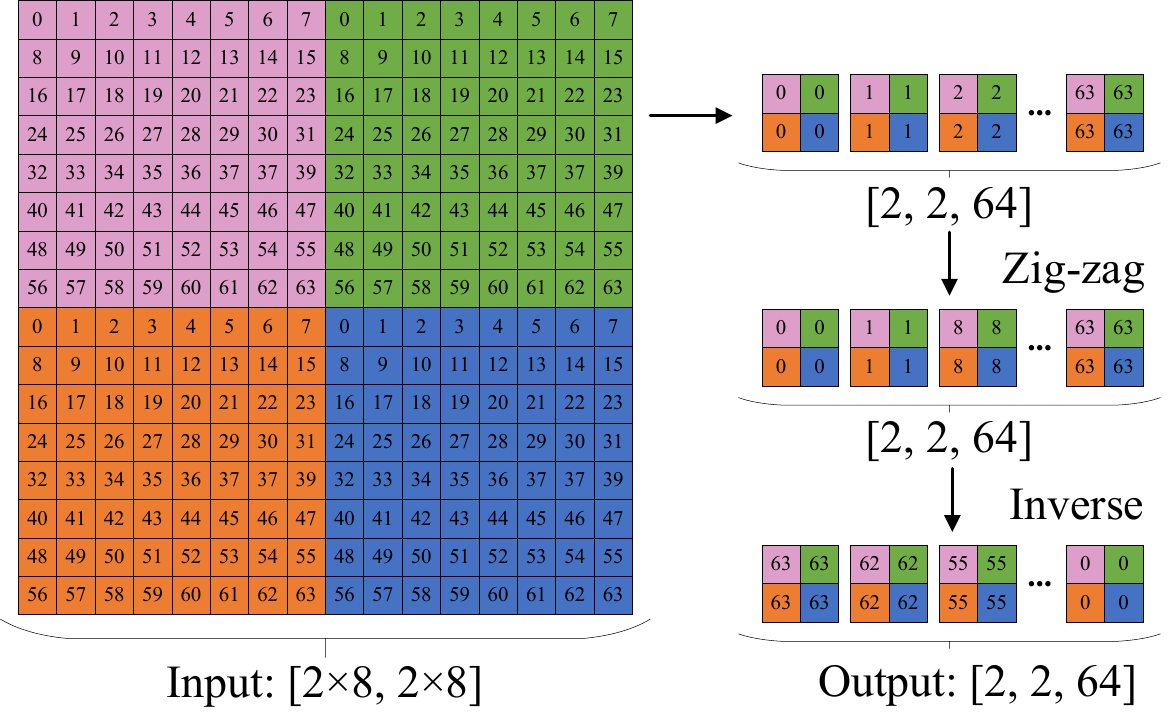}
  \caption{Data process. Using a $ 16 \times 16 $ image as an example, four $ 8 \times 8 $ DCT blocks are rearranged by frequency, zigzag scan and inverse order.}
  \label{fig:rearrange}
\end{figure}

\begin{figure}[htb]
  \centering
  \includegraphics[width = .25\textwidth]{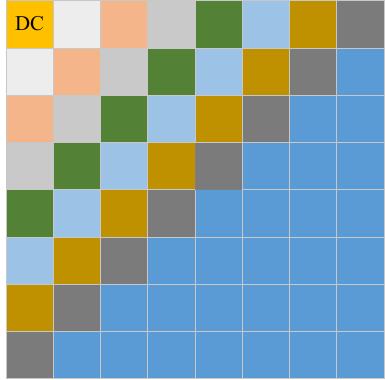}
  \caption{The corresponding frequency position for each column.}
  \label{fig:partition}
\end{figure}

\subsection{Similarity between frequencies}
Duda \cite{duda2020improving} points out that while DCT transform decorrelates data in a block, there remain other statistical dependencies like homoscedasticity. We all know that each coefficient in per $8\times 8$ block stands for one frequency. The top left corner of this block is the DC component, while the remaining 63 coefficients are AC components. 

To explore the similarity between different frequency coefficients, we  first take the coefficients of the same frequency in different blocks and form them into a vector, \eg frequency coefficients from $(h,  w, 8, 8)$ to $(h\times w, 8, 8)$. Next, we calculate the similarity of each frequency coefficient with the others. The results are shown in Figure~\ref{fig:in_block}. There are considerable similarities between low frequencies while less similarities between high frequencies. Moreover, the similarity of low frequency is scattered, and in most cases, frequencies located on its diagonal are strongly correlated.

As show in Figure~\ref{fig:rearrange}, we apply the same data processing as Guo~\etal~\cite{Guo_2022_CVPR} to rearrange each color component coefficients. And from the main text, the lengths of each column $\bm{y}^{(i)}_{j}$ are set as 28, 8, 7, 6, 5, 4, 3, 2 and 1 respectively. Therefore, our first column (\ie $\bm{y}_1^{(1:4)}$) in PPCM actually corresponds to 28 high frequency coefficients (blue girds in Figure~\ref{fig:partition}) while the last column (\ie $\bm{y}_9^{(1:4)}$) to the lowest frequency coefficient (\ie DC). Obviously, this division way is consistent with the distribution of frequency coefficients' similarities shown  in Figure~\ref{fig:in_block}, which demonstrates that why modeling context along these columns is effective to improve compression performance.

\section{More test details}
\label{sec:test}
\subsection{Test script}
In Table 3 mentioned in the main text, we compare throughput and latency with traditional codecs, including Lepton \cite{horn2017design}, JPEG XL \cite{alakuijala2019jpeg, alakuijala2020benchmarking} and CMIX \cite{cmix}. Here we provide more detailed test commands.

The throughput and latency of Lepton are obtained by its official benchmark tool. The command is as follows:
\begin{lstlisting}
#Lepton
#compression & decompression benchmark
lepton -benchmark -benchreps=32 -benchthreads=48 test.jpeg test.lep
\end{lstlisting}

The Benchmark tool provided by JPEG XL does not support JPEG recompression, so we use Python Multiprocessing \cite{multiprocessing} with 20 threads to test JPEG XL. 
The commands of compression and decompression are as follows:
\begin{lstlisting}
#jpeg xl 
#compressing
cjxl test.jpeg test.jxl --lossless_jpeg=1
#decompressing
djxl test.jxl test.jpeg
\end{lstlisting}

CMIX cannot set threads, so we use its default 1 thread. The command is as follows:

\begin{lstlisting}
#cmix 
#compressing
cmix -c test.jpeg test.cmix
#decompressing
cmix -d test.cmix test.jpeg
\end{lstlisting}

\subsection{More results of software optimization}
As shown in Tbale~3 in the main text, with proper software optimization, our fp16 model achieves comparable latency with int8 model, \ie 53 ms latency for compression and 54 ms latency for decompression on 1080p images. We  further  decompose the inference time of fp16 model in Table~\ref{tab:decomposition}. It shows model inference  consumes around 30\% latency while the entropy coder takes up over 50\% latency. 
Moreover, 
Table~\ref{tab:decomposition} demonstrates that there is still considerable space for further deployment optimization.

\begin{table}[]
    \centering
     \begin{tabular}{l|c|cl}
     	\hline
          &Phase & Time (ms)& \\
         \hline
         \multirow{4}{*}{Compress}
         &load image &6.52&12.1\% \\
         ~&Model Inference &16.33&30.3\% \\
         ~&Entropy encoder &29.80&55.3\%\\
         ~&Write bitstream &1.24&2.3\%\\
         ~&\textbf{Total} &\textbf{53.89}&\textbf{100\%} \\
         \hline
         \multirow{4}{*}{Decompress}
         &Decode prepare &1.08 &2.0\%\\
         ~&Model Inference &14.46 &26.7\%\\
         ~&Entropy decoder &29.48&54.4\%\\
         ~&Save image &9.18&16.9\% \\
         ~&\textbf{Total} &\textbf{54.20}&\textbf{100\%}\\
         \hline
    \end{tabular}
    \caption{Time decomposition for the whole compression or decompression process on 1080p images.}
    \label{tab:decomposition}
\end{table}

\section{More architecture details}
\label{sec:architectures}

\subsection{Hyper-network}
\label{sec:hyper-network}
\begin{figure}[htb]
  \centering
  \begin{subfigure}[b]{.9\linewidth}
      \centering
      \includegraphics[width=.99\linewidth]{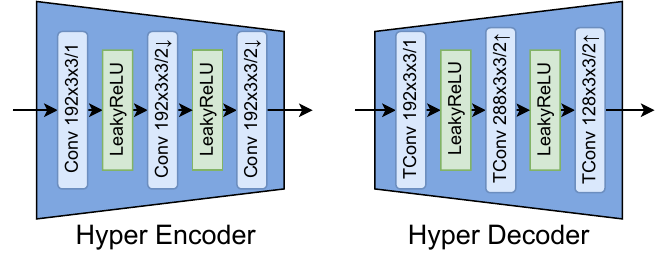}
      \subcaption{\textit{Hyper-network} in Y-Net.}
      \label{fig:y}
  \end{subfigure}
  \begin{subfigure}[b]{.9\linewidth}
      \centering
      \includegraphics[width=.99 \linewidth]{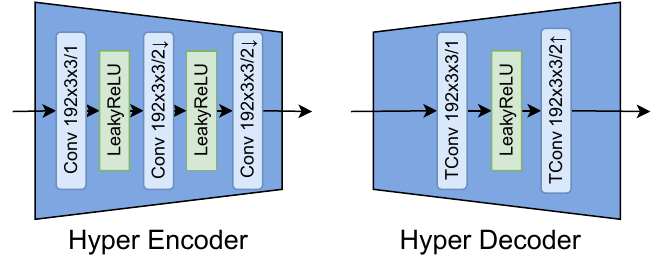}
      \subcaption{\textit{Hyper-network} in CbCr-Net.}
      \label{fig:cbcr}
  \end{subfigure}
  \caption{Detailed architectures of \textit{hyper-network} in Y-Net and CbCr-Net.}
  \label{fig:hyper}
\end{figure}

Figure~\ref{fig:hyper} shows the architecture details for \textit{hyper-network} mentioned in the main text.
Specially, we prune two layers of \textit{Hyper Decoder} in CbCr-Net and only upsample $2\times$, where the output of \textit{Hyper Decoder} (\ie $\bm{h}_{cbcr}$) has same resolution with $\bm{CbCr}_i$.

\subsection{Param-net}
\begin{figure}[htb]
  \centering
  \includegraphics[width=3cm]{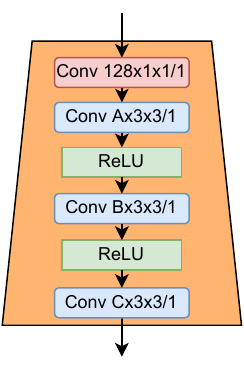}
  \caption{Parameter network (\textit{param-net}). $A$, $B$ and $C$ are decided by $ A = 128 - d, B = 128 - 2*d, C=2*n, d = (128 -2*n)//3$, where $n$ represent the channel number of next slice to be modeled}
  \label{fig:param}
\end{figure}

The detailed architecture of \textit{param-net} is shown in Figure~\ref{fig:param}. \textit{Param-net} takes $1\times 1$ convolution as the first layer to reduce the number of channels, which effectively saves the calculation.

\subsection{Full parallel PPCM}
\label{ssec:full_ppcm}
\begin{figure*}[htb]
  \centering
  \includegraphics[width=17cm]{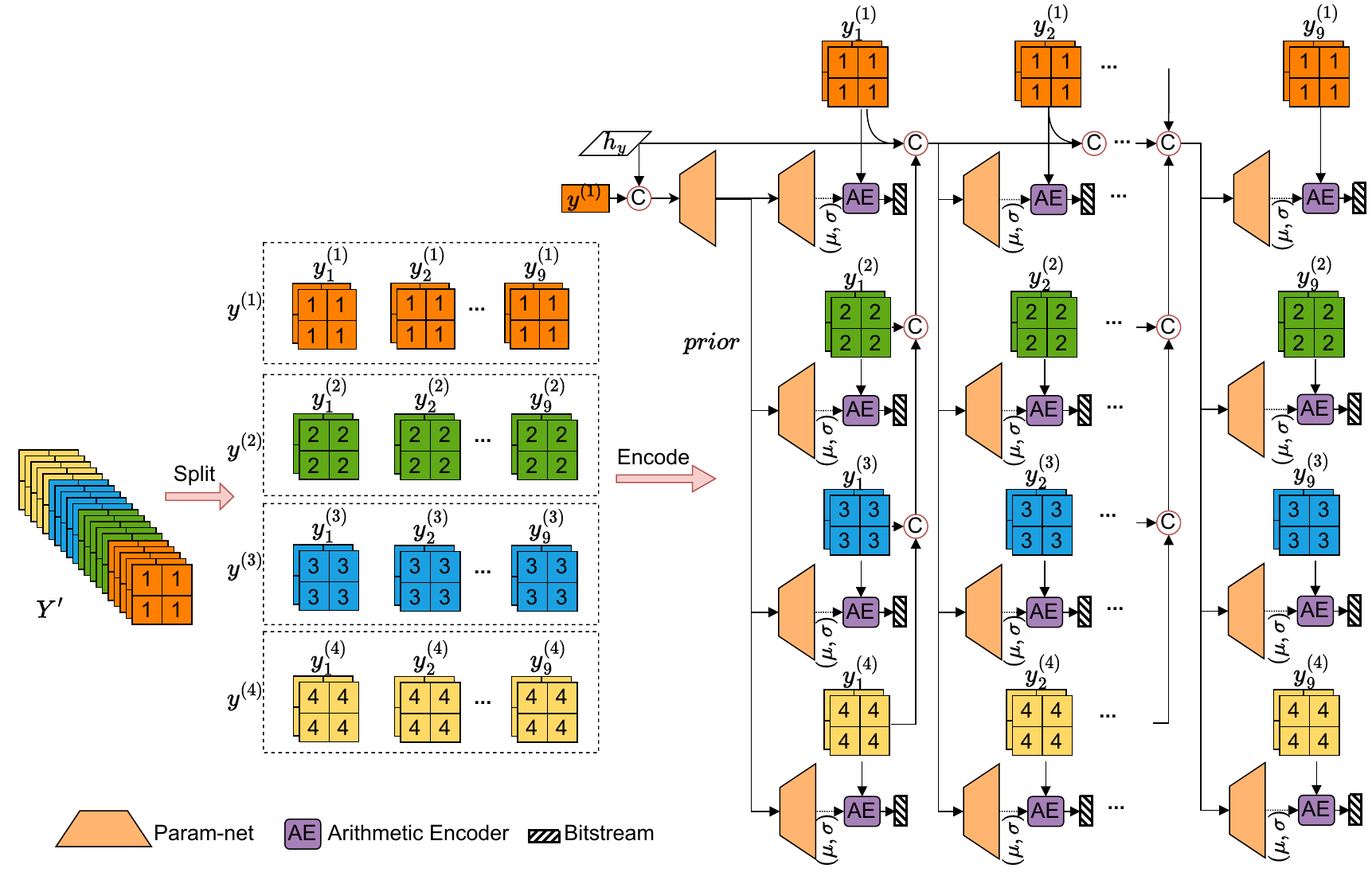}
  \caption{Compressing Y with full parallel PPCM and $\bm{h_y}$. Detailed architecture of \textit{param-net} is shown in Figure~\ref{fig:param}.}
  \label{fig:full_ppcm}
\end{figure*}

Figure~\ref{fig:full_ppcm} shows the architecture of the full parallel PPCM, which is a variant of PPCM mentioned in Section 4.\ 3 in the main text.

{\small
\bibliographystyle{ieee_fullname}
\bibliography{sup}
}